\documentclass[aps,twocolumn,floatfix,altaffilletter,superscriptaddress,preprintnumbers,
               tightenlines,showpacs,showkeys,notitlepage,nofootinbib]{revtex4-2}
\usepackage{hyperref}
\usepackage[normalem]{ulem}
\usepackage{amsmath,amssymb,slashed,cancel}
\usepackage[dvipsnames]{xcolor}
\usepackage{graphicx}
\usepackage{mathtools}
\usepackage{verbatim}
\usepackage{epsfig}
\usepackage{url}
\usepackage{multirow}
\usepackage{epstopdf}
\usepackage{siunitx}
\usepackage{enumitem}
\usepackage{cleveref}
\usepackage{lipsum}
\usepackage{gensymb}
\usepackage{titlesec} 
\allowdisplaybreaks

\makeatletter
\renewcommand{\thesection}{\arabic{section}}
\renewcommand{\thesubsection}{\thesection.\arabic{subsection}}
\renewcommand{\p@subsection}{}
\makeatother

\titleformat*{\section}{\centering\bfseries}
\titlelabel{\thetitle\quad}
\titleformat*{\paragraph}{\bfseries}
\titlespacing*{\paragraph}{0pt}{3.25ex plus 1ex minus .2ex}{1em}

\makeatletter
\def\l@subsubsection#1#2{}
\makeatother

\definecolor{rossoferrari}{HTML}{D9073D}
\definecolor{mediumblue}{HTML}{0000CD}
\definecolor{forestgreen}{HTML}{228B22}
\definecolor{desy_blue}{HTML}{009EE2}
\definecolor{desy_orange}{HTML}{FD8800}
\definecolor{light_pink}{rgb}{1,0.4,0.4}
\definecolor{light_blue}{rgb}{0.284602,0.317763,0.963947}

\hypersetup{
setpagesize=false,
bookmarksnumbered=true,%
bookmarksopen=true,%
colorlinks=true,%
linkcolor=light_blue,
urlcolor=blue,
citecolor=rossoferrari,
linktocpage=false,
}

\newcommand{\ev}[1]{\ensuremath{\left\langle #1 %
                     \right\rangle}} 

\newcommand{\diag}{\text{diag}}

\renewcommand{\vec}[1]{{\mathbf{#1}}}

\newcommand{\be}{\begin{equation}}
\newcommand{\ee}{\end{equation}}
\newcommand{\bea}{\begin{eqnarray}}
\newcommand{\eea}{\end{eqnarray}}

\usepackage{soul}

\usepackage[nolist]{acronym}
\begin{acronym}
	\acro{SM}{Standard Model}
	\acro{BSM}{Beyond the Standard Model}
	\acro{NP}{New Physics}
    \acro{GW}{Gravitational Wave}
    \acro{GWs}{Gravitational Waves}
    \acro{HF}{High Frequency}
\end{acronym}
\begin{document}

\title{High-Frequency Gravitational Wave Detection \\
       via Optical Frequency Modulation}

\author{Torsten Bringmann}
\affiliation{Department of Physics, University of Oslo, 0316 Oslo, Norway}
\affiliation{Theoretical Physics Department, CERN, 1211 Geneva 23, Switzerland}

\author{Valerie Domcke}
\affiliation{Theoretical Physics Department, CERN, 1211 Geneva 23, Switzerland}

\author{Elina Fuchs}
\affiliation{Theoretical Physics Department, CERN, 1211 Geneva 23, Switzerland}
\affiliation{Institut f\"ur Theoretische Physik, Leibniz Universit\"at Hannover, 30167 Hannover, Germany}
\affiliation{Physikalisch-Technische Bundesanstalt, 38116 Braunschweig, Germany}

\author{Joachim Kopp}
\affiliation{Theoretical Physics Department, CERN, 1211 Geneva 23, Switzerland}
\affiliation{PRISMA+ Cluster of Excellence \& Mainz Institute for Theoretical Physics,
             55128 Mainz, Germany}

\preprint{CERN-TH-2023-065, MITP-23-017}



\begin{abstract}
High-frequency gravitational waves can be detected by observing the frequency modulation they impart on photons.  We discuss fundamental limitations to this method related to the fact that it is impossible to construct a perfectly rigid detector.  We then propose several novel methods to search for $\mathcal{O}(\si{MHz}-\si{GHz})$ gravitational waves based on the frequency modulation induced in the spectrum of an intense laser beam, by applying optical frequency demodulation techniques, or by using optical atomic clock technology. We find promising sensitivities across a broad frequency range.
\end{abstract}

\maketitle

\paragraph*{Introduction}
Our Universe is filled with gravitational waves (GWs)
which render space and time themselves highly non-static.
Photons traveling through such an
environment are affected by GW-induced spacetime ripples in manifold 
ways, reminiscent of the way a water craft is affected by rough seas.

Here we focus on modulations of the photon frequency
which arise due to variations of 
the gravitational field along the photon trajectory and due to boundary conditions imposed by the photon emitter and 
absorber, such as Doppler shift.
The goals of this letter are twofold: first, we discuss the physics underlying GW-induced phton frequency modulation and calculate its magnitude, with a focus on the distinction between detectors composed of free-falling test masses and detectors that are rigid.
While we find that in the latter case the sensitivity grows as $\omega_g L$ (where $\omega_g$ is the angular frequency of the 
GW and $L$ is the size of the detector), we demonstrate that this effect is spurious in the limit of large $\omega_g$.
In the second part of this work we 
propose several promising new methods for searching for high-frequency GWs, based on experimental 
methods from 
quantum optics: {\it (i)} detection of sidebands
in the spectrum of an 
intense laser; {\it (ii)} optical frequency demodulation to convert frequency shifts into an amplified electrical signal; 
{\it (iii)} an 
``optical rectifier'' to ensure that the detected photons have a non-zero net frequency shift which can be measured using atomic clock techniques.

The impact of GWs on photons has previously been studied in 
refs.~\cite{Kaufmann:1970, Estabrook:1975, Tinto:1998ee, Kopeikin:1998ts, Tinto:2002de, Siparov, Armstrong:2006, Lobato:2021ffr, Weinberg:2008zzc}, 
while using optical atomic clock technology to search for GWs has been proposed in 
refs.~\cite{Loeb:2015ffa, Vutha:2015aza, Kolkowitz:2016wyg}, 
albeit for much lower frequencies.
\\[-5ex]


\paragraph*{Photon frequency shift.}
We are interested in comparing the frequency of a photon, $\omega_\gamma$, as measured by two different observers which we will denote source ($S$) and detector ($D$), respectively.
We define the origin of our coordinate system to be the spacetime point at which the photon is emitted.
We assume $D$ is placed on the positive $x^1$-axis, and that the photon has initial 4-momentum $p^\mu \big|_{t=0} = (\omega_0,\omega_0,0,0)$ in the frame of a free-falling observer.
In a frame with metric $g_{\mu\nu}$, an observer moving with four-velocity $u^\mu$ will measure a photon frequency
$\omega_\gamma = -g_{\mu\nu} p^\mu u^\nu$.

Here we want to investigate possible effects due to tiny space-time perturbations induced by a GW that passes through $S$ and $D$ initially at rest. We write
\begin{align}
    g_{\mu\nu} &= \eta_{\mu\nu}+h_{\mu\nu} \\
    p^\mu      &= (\omega_0,\omega_0,0,0) + \delta p^\mu \\
    u^\mu      &= (1,0,0,0) + \delta u^\mu \,,
\end{align}
with $\eta_{\mu\nu} = \diag(-1,1,1,1)$, and with $h_{\mu\nu}$, $\delta p^\mu$, and $\delta u^\mu$ denoting $\mathcal{O}(h)$ corrections to the corresponding quantities, where $h$ is the GW amplitude (``strain'').
We thus obtain
\begin{align}
    \omega_\gamma = \omega_0 (1 \!+\! \delta u^0 \!-\! \delta u^1 \!-\! h_{00} \!-\! h_{01})
                  + \delta p^0 + \mathcal{O}(h^2) \,,
    \label{eq:omega}
\end{align}
where $p^0$ obeys
the geodesic equation
\begin{align}
    \frac{dp^0}{d\lambda} \!=\! -\Gamma^0_{\mu\nu} p^\mu p^\nu
                          = -\omega_0^2 \big( \Gamma^0_{00} \!+\! 2\Gamma^0_{10}
                                            \!+\! \Gamma^0_{11} \big) + \mathcal{O}(h^2) \,.
\end{align}
Here, $\Gamma^\rho_{\mu\nu}$ denote the Christoffel symbols and $\lambda$ is the affine parameter 
that parameterizes the photon geodesic, with $\lambda = 0$ corresponding to $t=0$.
At leading order in $h$, it is sufficient to  evaluate the Christoffel symbols at  $x^\mu = x^\mu_{\lambda,0} \equiv (\lambda\omega_0,\lambda\omega_0,0,0)$. Hence, we find
\begin{align}
    \delta p^0 = -\omega_0^2 \int_0^{\lambda_D} \! d\lambda' \big[ \Gamma^0_{00}
                     + 2 \Gamma^0_{10} + \Gamma^0_{11} \big]_{x^\mu=x^\mu_{\lambda',0}} \,,
\end{align}
with $\lambda_D$ being the value of $\lambda$ at the spacetime point where the photon is detected.
Plugging the above expression into \cref{eq:omega}, and performing some algebra,  we arrive at our master formula for the observed frequency shift,
\begin{align}
    \frac{\omega_\gamma^D - \omega_\gamma^S}{\omega_\gamma^D}
        &= -\frac{\omega_0}{2} \int_0^{\lambda_D} \!\! d\lambda'\,
            \partial_0 \big[ h_{00} \!+\! 2h_{10} \!+\! h_{11} \big]_{x^\mu=x^\mu_{\lambda',0}}
                                                                  \nonumber\\[0.2cm]
        &  \quad + \Big[ \delta u^0 - \delta u^1 \Big] (\lambda_D) 
             - \Big[ \delta u^0 - \delta u^1 \Big] (\lambda_S)  \,.
    \label{eq:fshift}   
\end{align}
Here, $\omega_\gamma^S$ is the frequency with which the photon is emitted by the source $S$ at $t=0$, and 
$\omega_\gamma^D$ is its frequency as measured by $D$.
Let us stress that this result is fully general, and in fact valid for any weak gravitational field. 
The terms in the first line describe the effect of a varying gravitational field along the entire
photon trajectory.
The terms in the second line describe additional effects due to source and detector motion.
\\[-5ex]


\paragraph*{Free-falling detectors.}
We start with the case of $S$ and $D$ being in free fall. (In practice, one only needs to require that they can move freely in the direction of photon propagation.)
This situation is most easily described in the  transverse--traceless (TT) gauge, defined by the conditions 
\begin{align}
    h^{TT}_{\mu0} = 0 \,,
    \qquad
    \partial^i h^{TT}_{ij} = 0 \,,
    \qquad
    \eta^{ij}h^{TT}_{ij}=0 \,.
\end{align}
In this gauge, observers initially at rest remain at rest, and hence $\delta u^0 = \delta u^i = 0$~\cite{Maggiore:2007ulw}.
Also the metric perturbation takes a particularly simple form, with
\begin{align}
\label{eq:htt11}
    h^{TT}_{11}(x^\mu) = h_+ s^2_\vartheta \, \cos\!\left[ \omega_g (x^0 - c_\vartheta x^1
                                                         - s_\vartheta x^3) + \varphi_0 \right]
\end{align}
for a plane GW propagating in the $x^1$--$x^3$ plane at an angle $\vartheta$ from the $x^1$ axis.
Here, $\varphi_0$ is the GW phase at $\mathbf{x}=\mathbf{0}$ when the photon is emitted, and $s_\vartheta \equiv \sin\vartheta$ and $c_\vartheta \equiv \cos\vartheta$.
The GW strain is denoted by $h_+$, with the `+' sign indicating a polarization where one of the two quadrupole axes is aligned with the $x^2$ axis. 
GWs with the orthogonal (`$\times$') polarization
will not affect photons propagating along the $x^1$ axis.
From \cref{eq:fshift}, we directly obtain
\begin{align}
    \frac{\omega_\gamma^D - \omega_\gamma^S}{\omega_\gamma^D}
        &= h_+ c^2_{\vartheta\!/\!2}             
        \Big\{\!
                 \cos \varphi_0- \cos\!\big[\omega_g L (1 \!-\! c_\vartheta) + \varphi_0 \big]
            \Big\} ,
    \label{eq:fshift_freefall}
\end{align}
consistent with previous results~\cite{Estabrook:1975, Lobato:2021ffr, Kolkowitz:2016wyg}. Here, $L = x_D^0 + \mathcal{O}(h)$ is the coordinate distance between the photon source and detector.

It is instructive to repeat the same calculation in the \emph{proper detector frame}, where spatial coordinates are defined by the 
distances an observer using a rigid ruler would measure  \cite{Maggiore:2007ulw}.
While a source placed at the origin will remain at rest also in this frame, $x_D$ changes with time due to the force exerted by 
the GW.
As expected and explicitly demonstrated in the Appendix,
deriving \cref{eq:fshift_freefall} in the proper detector 
frame gives the same result as in the TT frame.
However, this calculation reveals an intriguing cancellation between terms coming from the first and second line in \cref{eq:fshift}.
Remarkably, the additional terms that eventually cancel generically grow with $\omega_g L$, that is, if the cancellation could be
avoided, they would lead to strongly enhanced experimental sensitivities for $\omega_g\gg 1/L$.
\\[-5ex]


\paragraph*{Rigid experimental setups.}
Partially motivated by this observation, we next consider a situation where $S$ and $D$ are not free-falling, but are kept at rest in 
the proper detector frame, i.e.~$\delta u^0 = \delta u^i = 0$.
This leads to a frequency shift
that contains terms growing with $\omega_g L$. In the most favourable case of an incoming
GW perpendicular to the laser beam, $\vartheta=\pi/2$,
the general result from \cref{eq:photonprop} in the Appendix simplifies to 
\begin{align}
    \label{eq:fshift_PD} 
    \frac{\omega_\gamma^D - \omega_\gamma^S}{\omega_\gamma^D}
        &= \frac{h_+}{2}\Bigg\{
               \cos\varphi_0 - \omega_g L  \sin(\omega_g L + \varphi_0) \\
        &\qquad\quad
             +\left(\frac12\omega_g^2 L^2-1 \right)  \cos(\omega_g L + \varphi_0)
           \Bigg\}\,.\notag
\end{align}
At face value this would imply detection prospects that are even enhanced as $\sim(\omega_g L)^2$ at high 
GW frequencies.
To the experienced reader, this may sound too good to be true.
And indeed, constructing a perfectly rigid ruler on such scales turns out to 
be impossible. To show this, we model the material separating the photon source and detector as a chain of harmonic 
oscillators in the $x^1$-direction.
Writing the displacement of an oscillator at $x^1$ from its rest position as $\xi$, 
the oscillator equation for such a system is \cite{Maggiore:2007ulw}
\begin{align}
    \ddot \xi - \frac{\omega_0^2 L^2}{\pi^2} \xi'' + \gamma \dot \xi
        = \frac{1}{2} x^1 \ddot{h}_{11}^{TT} \,,
    \label{eq:oscillator}
\end{align}
where $\ddot{\xi} \equiv d^2\xi/dt^2$, $\xi'' \equiv d^2\xi/d(x^1)^2$, $\omega_0$ is the resonance frequency of the fundamental 
mode of the system, $\gamma$ is a damping coefficient, and we have again used $\vartheta = \pi/2$.
Note that \cref{eq:oscillator} is the equation of motion of an oscillator in the proper detector frame, even though the metric 
perturbation is the much simpler one from the TT frame.
In the limit $\omega_g \gg \omega_0, \gamma$, the last two terms on the left hand side become negligible, 
since $\xi \propto h L$, $\dot \xi \propto \omega_g h L$, $\ddot \xi \propto \omega_g^2 h L$, $\xi'' \propto h / L$ and 
$\ddot h_{11}^{TT} \sim \omega_g^2 h$.
In this limit, the equation therefore becomes identical to that of a free-falling test mass in the PD frame, 
cf.\ \cref{eq:du-dtau,eq:Christoffel,eq:h00PD,eq:h01PD},
with correction terms suppressed by $\omega_0^2 / \omega_g^2$.
At large $\omega_g$, the two extremities of our `rigid' setup are thus responding to the GW just like free-falling 
test masses would, which is quite the opposite of what one would expect from a rigid system.
(In the opposite limit $\omega_g \ll \omega_0$, on the other hand, we recover the expected behaviour of a rigid 
ruler~\cite{Maggiore:2007ulw}.)

The fact that \cref{eq:fshift_PD} cannot be na\"{i}vely applied for $\omega_g L \gtrsim v_s$, 
where $v_s$ is the velocity of sound in the detector, 
is by itself an important observation for the construction of high-frequency GW detectors and one of the main results of this letter.
Our explicit calculations above assume a very simple setup, i.e.\ a source and detector separated by some material. Similar 
arguments will however also apply to more complex systems, such as electronic equipment used to generate `static' 
electromagnetic fields in electromagnetic GW detectors~\cite{Berlin:2021txa, Domcke:2022rgu}.
Moreover, even though using the idealized setup of a rigid ruler, our discussion illustrates the general 
importance of the boundary conditions in \cref{eq:fshift}.
Suitable choices of material and suspension thus have the potential of influencing the sensitivity of high 
frequency GW searches, and need to be studied carefully on a case-by-case basis.
\\[-5ex]


\paragraph*{Frequency modulation of a laser beam.}
We now consider a continuous flux of photons in a laser beam.
In \cref{eq:fshift_freefall,eq:fshift_PD}, the photon frequency is then
modulated by the phase $\varphi_0(t)=\omega_g (t-x)$  
of the GW at the time of photon emission, where (without loss of generality) we have set the phase to zero at $(t,x) = (0,0)$. 
As we observe the photons arrive at the detector over some finite time interval, $\varphi_0$
oscillates with frequency $\omega_g$.
For a photon coherence length $\gg 1/\omega_g$, this leads to sidebands at 
$\omega_\gamma^\pm \equiv \omega_\gamma^S \pm \omega_g$ 
in the spectrum.
Quantitatively,  the emitted photon wave takes the form $A(t, 0) = A_\gamma \cos(\omega_\gamma^S t + \phi_\gamma)$,
with amplitude $A_\gamma$ and phase $\phi_\gamma$.
For $\omega_g \ll \omega_\gamma$, we can write for the photon wave at the position of the detector\footnote{
We give a more rigorous derivation of this result for arbitrary values of $\vartheta$ in App.~\ref{app:MW}, based on solving Maxwell's equations in curved space time.
}
\begin{align}
  \frac{A(t, L)}{A_\gamma} &=\cos\left(\int_0^t \omega_\gamma^D(t') dt' + \phi'_\gamma \right)
   = \cos(\omega^S_\gamma t + \phi'_\gamma ) \notag\\
     &\hspace{-0.7cm} + \frac{h_+}{4} \frac{\omega^S_\gamma}{\omega_g} \Big[ \cos[\omega^+ (t-L) + \phi'_\gamma] - \cos[\omega^- (t-L) + \phi'_\gamma]  \notag \\
     &\hspace{-0.7cm}   - \cos[\omega^+ t - \omega_\gamma L + \phi'_\gamma] +  \cos[\omega^- t - \omega_\gamma L + \phi'_\gamma]  \notag \\
    &\hspace{-0.7cm} - 2 \sin[\omega_g L] \sin[\omega_\gamma (t-L) + \phi'_\gamma] \Big],
     \label{eq:A}
\end{align}  
where we denote 
with $\phi'_\gamma$ the  photon phase at $t = 0$ and $x = L$.
For simplicity, we have assumed here and in 
the following that source and detector are freely falling, as in \cref{eq:fshift_freefall,eq:fshift_PD},
and that $\vartheta = \pi/2$. 
The first term on the right-hand side of \cref{eq:A} is the carrier wave, i.e.~the unperturbed photon signal.
The last term describes a tiny correction to the amplitude of the carrier wave; this
is irrelevant in practice. The remaining four terms generate the sidebands.
In the following, we discuss three different ways that may allow the detection of such a signal.
\\[-5ex]


\paragraph*{Direct observation of sidebands.}
For large $\omega_g$, the sidebands in \cref{eq:A} are separated by a relatively large frequency gap from the 
carrier frequency $\omega^S_\gamma$. However, their {\it intensity} is suppressed by $h_+^2$.
(Experimental attempts to detect the interference term, linear in $h_+$, would have to deal with an overwhelming 
background of photons from the main carrier line.
Heterodyne detection schemes,
modulating the carrier line with a beat frequency such that the frequency difference between this beat frequency and the GW frequency becomes tractable for readout,
may provide an interesting possibility to overcome this challenge and are left for future work.)  
Detecting such faint sidebands requires a powerful photon source that is
highly monochromatic, complemented by 
a very efficient optical filter system that removes the carrier frequency after the photons have propagated 
to the detector.

\begin{figure}
    \centering
    \includegraphics[width=\columnwidth]{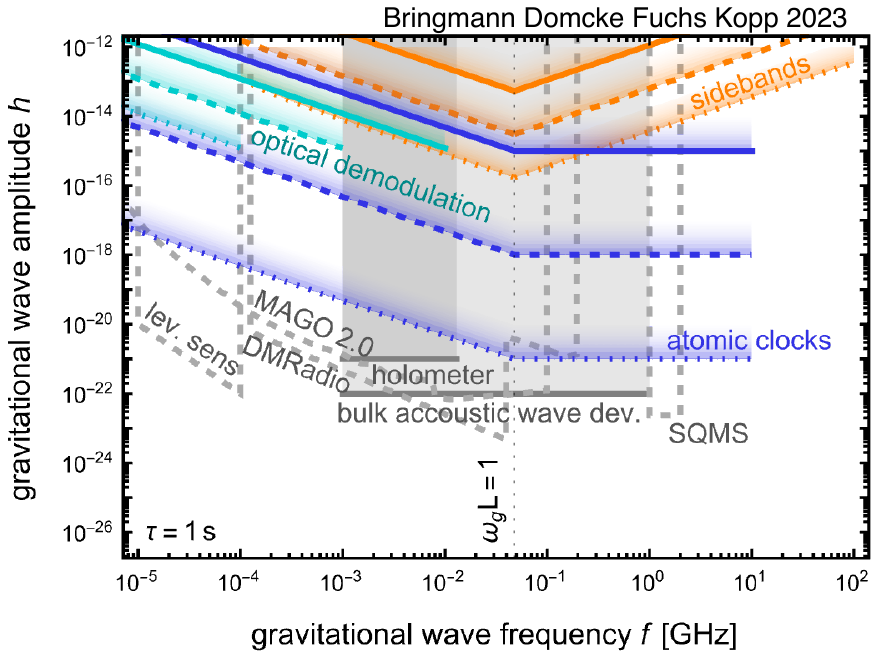}
    \caption{Sensitivity estimates for the three novel high-frequency GW detection methods proposed here (colored lines).
    For each proposal we show the sensitivity under conservative (solid), realistic (dashed), and optimistic (dotted) assumptions for the achievable experimental sensitivity, in particular
    $(\alpha_\text{T}, \alpha_\text{th}) = \{ (10^{-10}, 10^{-15}), (10^{-15}, 10^{-17}), (10^{-20}, 10^{-19}) \}$,
     $\sigma = 10,\ 1,\ 0.1$\;MHz and $\delta = 10^{-15}, 10^{-18}, 10^{-21}$. 
    In all cases we have set $\tau=1$\,s, $L = \SI{1}{m}$, $\omega_\gamma^S /2 \pi = \SI{2e14}{Hz}$ and $P = \si{mW}$.
    The grey shaded regions indicate other existing (solid) and proposed (dashed) experiments, in particular levitated sensors~\cite{Aggarwal:2020umq}, axion haloscopes such as DMRadio GUT~\cite{Domcke:2022rgu}, the holometer experiment~\cite{Holometer:2016qoh}, bulk acoustic wave devices~\cite{Goryachev:2014yra} and microwave cavities like SQMs~\cite{Berlin:2021txa} and MAGO~2.0 \cite{Berlin:2023grv}.
    \label{fig:sens}}
\end{figure}

In this respect, optical cavities or techniques from fiber optics may offer a promising avenue towards table-top high-frequency 
GW detectors. Let us consider a filter of width $\Delta \lambda$ which suppresses the main carrier frequency by 
$\alpha_T \ll 1$ while ensuring an ${\cal O}(1)$ transmission  at the location of the sideband. 
In \cref{fig:sens} we consider filter efficiencies of $\alpha_T = 10^{-10}..10^{-20}$ and a bandwidth of $\Delta \lambda \simeq \SI{100}{kHz}$, which may e.g.\ be achieved by employing optical cavities tuned to the sideband frequency~\cite{Matei:2017kug}, or potentially also by stacking multiple fiber Bragg gratings~\cite{Canning:2008, Chen:2011}.
We will neglect propagation effects induced by the GW in this filtering system, noting that they can be suppressed by choosing a suitable geometry (e.g.\ parallel to the incoming GW).

We will further assume that the sensitivity to a gravitational 
wave signal is only limited by the requirement to find a sufficiently large number $s$ of signal photons in the side bands.
From \cref{eq:A}, we have
$s \simeq ({P \, \tau}/{\omega_\gamma^S}) \times h_+^2 ({\omega_\gamma^S}/{\omega_g})^2 \; \text{min}(1, \omega_g^2 L^2)$,
where $P$ is the laser power and $\tau$ is the signal duration or measurement time, whichever is shorter.
The sensitivity curves we show as orange lines in \cref{fig:sens} assume a mW laser  emitting at a wave length of \SI{1500}{nm}. 
They are based on requiring $s$ to be larger than the square root of the number of spillover photons from the carrier 
mode, $n_\text{s.o.} \simeq \alpha_T P \, \tau / \omega_\gamma^S$ plus the number of photons due to thermal noise in the 
optical filter system, 
$n_\text{th} \simeq p^\text{th} (\omega_\gamma^S + \omega_g, \alpha_\text{th}) \, P \, \tau / \omega_\gamma^S$. 
Here, we take the thermal noise spectrum $p^\text{th}(\omega, \alpha_\text{th})$ to be a Lorentzian centered at 
$\omega$ and with relative width $\alpha_\text{th}$.
Optical fiber links with noise levels below $\alpha_\text{th} \simeq 10^{-17}$ have been described in 
refs.~\cite{Akatsuka_2014, Schioppo:2022iqe}.
\\[-5ex]


\paragraph*{Optical demodulation.}
A GW modulates the frequency of a propagating photon in the same way an FM radio 
transmitter modulates the frequency of an FM carrier, for which advanced demodulation techniques 
exist. Inspired by refs.~\cite{Barot:2020, Barot:2022}, we propose here to split the incoming photon beam and filter it 
through two slightly detuned optical cavities
(alternatively, a setup using fiber Bragg gratings could be envisioned).
The system is adjusted such that the carrier frequency lies exactly between the transmission peaks of the two cavities,
and the two filtered components interfere destructively in a photon detector.
A slight frequency shift due to the modulation will then enhance the signal in one cavity with respect to the other, 
disrupting the destructive interference and thus creating a non-zero signal in the detector.
This signal is converted into a voltage whose evolution with time carries the information originally encoded in the 
optical wave.

The main challenge is the tiny $\mathcal{O}(h)$ amplitude of the frequency modulation.
Notably, the relative width of the carrier mode should be smaller than $h$
-- in which case the Heisenberg uncertainty principle dictates coherence times much longer than $2\pi/\omega_g$. 
%
Another relevant consideration is laser shot noise.
The degree to which destructive interference between the two filtered beams can be realized is subject to 
Poisson fluctuations in the intensity of each mode.
Assuming transmission profiles shifted by half of the profile width $\sigma$ to either side of the carrier frequency, we estimate 
the number of signal photons after interference as $s \simeq h_+ P \tau \, \text{min}(\omega_g L, 1) / \sigma$.
%
Notably, the number of signal photons now scales linearly and no longer quadratically with the GW amplitude.
Requiring $s$ to be larger than the square root of the shot noise, $n_\text{sh} \simeq P \tau / \omega_\gamma^S$, yields the 
sensitivity estimates shown in \cref{fig:sens} (cyan lines).
In the plot, we have considered values of $\sigma$ between \SI{10}{MHz} and \SI{0.1}{MHz}.
In this case thermal noise is subdominant as long as $\alpha_\text{th} \omega_\gamma^S \ll \sigma$.
Note that small $\sigma$ implies long integration times (high finesse);
for $\omega_g \gtrsim \sigma$, the photons' retention time inside the cavity would be $> 1/\omega_g$, and the signal would average to zero, as indicated by the endpoint of the cyan lines in \cref{fig:sens}.
\\[-5ex]


\paragraph*{Atomic clock techniques.}
The most powerful methods for detecting tiny frequency shifts in optical signals have been developed in the context of optical 
atomic clocks \cite{Ludlow:2015, Safronova:2017xyt}, 
allowing e.g.~for gravitational redshift measurements over $\mathcal{O}(\si{mm})$ distances \cite{Bothwell:2021fqe}.
The techniques of refs.~\cite{Loeb:2015ffa, Vutha:2015aza, Kolkowitz:2016wyg},  however, based on optical lattice clocks in a 
space-based GW detector, are not directly applicable for our purposes:
achieving the desired frequency resolution would require extremely narrow optical lines, and hence 
long integration times beyond several seconds.
The effect of a high-frequency GW would average out over such time intervals.

Here, we propose to use an ``optical rectifier'' to prevent averaging.
In the simplest case, this means blocking the optical signal during half of each GW period using a shutter.
The net frequency shift of the photons will then be non-zero, 
$\Delta \omega_\gamma \sim h_+ \omega_\gamma^S \, \text{min}(1, \omega_g L)$,
and can be detected for instance by 
comparing the shifted photon frequency to an atomic reference transition using Ramsey spectroscopy \cite{Kolkowitz:2016wyg}.
By splitting off part of the beam before the optical rectifier, the laser can be locked to the atomic transition such that 
a passing GW appears as de-tuning of the beam passing the optical rectifier.
Such an approach seemingly requires prior knowledge of the GW frequency and phase, 
which is of course not possible.
However, if the GW signal features a broad, or time-varying, frequency spectrum, 
being sensitive in only a very narrow frequency interval is sufficient to make a detection.
We propose to operate at least two detectors in parallel, with the phases of their optical rectifiers offset by $\pi/2$, such that at least one of them can observe a non-zero signal.

The resulting GW sensitivity shown in \cref{fig:sens} (blue lines) is based on a sensitivity to frequency shifts of
$\Delta \omega_\gamma/\omega_\gamma^S \simeq 
\delta \cdot (\SI{1}{sec} / \tau)^{1/2}$,
assuming
that the measurement is limited by statistical uncertainty. 
Here, as before, $\Delta \omega_\gamma$ is given by eq.~\eqref{eq:fshift_freefall}.
Accuracies of $\delta \simeq {\{\num{3 e-18}, \num{9.7 e-18}\}}$ have been achieved with optical clock techniques in refs.~\cite{Bothwell:2021fqe,Zheng:2021rnx}, respectively.
With precision doubling roughly every year \cite{Safronova:2017xyt, Arias:2018}, significant future improvements seem possible. 
$^{229}$Th nuclear clocks~\cite{Campbell:2012, Peik:2020cwm} are expected to reach even better precision.
The method outlined here is limited at low frequencies by the stability of the laser and at high frequencies by the shutter speed, but we expect it to be feasible within the frequency range shown in \cref{fig:sens}. 
\\[-5ex]


\paragraph*{Comparison with other limits.}
Gray lines in \cref{fig:sens} show existing (solid, shaded) and projected (dashed) limits on high-frequency GWs from the literature.
We caution that these serve to guide the eye only, due to different search strategies and assumptions on the signal duration.
Concretely, for references~\cite{Holometer:2016qoh, Goryachev:2014yra, Aggarwal:2020umq} quoting bounds or projected 
sensitivities in terms of a power spectral density, $(S_n)^{1/2}$, 
we show the limit on the amplitude of 
coherent signals obtained as $h < (S_n/\tau)^{1/2}$ whereas for references~\cite{Berlin:2021txa, Domcke:2022rgu} 
quoting bounds on the dimensionless gravitational 
wave amplitude based on an observation time $T_\text{obs}$, we rescale 
the limits as $(T_\text{obs}/\tau)^{1/4}$, assuming the coherence of the signal to be limited only by the signal duration.
Finally, for \cite{Berlin:2023grv} we show the limits on the GW amplitude as quoted therein since with the proposed scanning strategy, the sensitivity is limited by the measurement time per frequency and not the signal duration in most of the parameter space of interest.
For recasting~\cite{Domcke:2022rgu} we have assumed an observation time per frequency bin $T_\text{obs}$ of 1~s, for \cite{Berlin:2021txa,Berlin:2023grv} we work with the values quoted in those references. 
In \cref{fig:sens} we adopted a common signal duration of $\tau = \SI{1}{sec}$ (in \cref{fig:sens2} in the Appendix we show, for  
comparison, the case of $\tau = \SI{e5}{sec}$). See ref.~\cite{Aggarwal:2020olq} for details.
\\[-5ex]

\paragraph*{Conclusions.} 
We revisited the frequency modulation of photons in a GW background, pointing out fundamental
limits to detecting this effect and proposing three novel experimental setups which promise
highly competitive sensitivities to high-frequency GWs.
We stress that the methods outlined here are in no way expected to be exhaustive of the 
possibilities of searching for GW induced frequency shifts in optical systems.
In fact, our work aims to trigger
more in-depth studies of these and related ideas.
To further aid this development, we summarize in the Appendix some of our `failed attempts' of using 
electromagnetic precision experiments in this context, hoping that the lessons learned from these considerations
might be instructive.


\paragraph*{Acknowledgments.}
It is a pleasure to thank Wolfram Ratzinger for very illuminating discussions on the Doppler shift of photons in a GW background, 
Johannes Skaar for educating us on Bragg filters and fiber optics in general,
Jun Ye for innumerable crucial insights into the physics of atomic clocks, 
Fritz Wagner for sharing his expertise on the M\"ossbauer effect,
Klemens Hammerer for useful explanations on sideband detection with cavities,
Tom Melia, Piet Schmidt, and Tadahiro Takahashi for illuminating discussions on optical clock comparisons,
Lingze Duan for important insights on optical demodulation,
Camilo Garcia Cely, Sebastian Ellis, Sung Mook Lee and Nick Rodd for their insights on comparing high-frequency GW sensitivities as well as on the quirks of the proper detector frame,
and Clara Murgui for her explanations around optomechanical cavities.
We moreover thank Camilo Garcia Cely, Lingze Duan, Klemens Hammerer, and Tadahiro Takahashi for valuable comments on the manuscript.
E.F.\ acknowledges support by the Deutsche Forschungsgemeinschaft (DFG, German Research Foundation) under Germany’s Excellence Strategy –– EXC-2123 “QuantumFrontiers”
— 390837967.


\vspace*{.7cm}
\begin{center}
    \bfseries APPENDIX
\end{center}
\vspace*{-.7cm}

\appendix
\section{The proper detector frame}
\label{sec:PDframe}

Our goal here is to evaluate \cref{eq:fshift} in the main text for free-falling setups, but working in the proper detector (PD) frame, 
where spatial coordinates $x^i$ are defined by what an observer using a rigid ruler would measure.
We expect to reproduce the TT frame result from \cref{eq:fshift_freefall}.

Let us first recall
that in order to compute the effect of the changing gravitational potential on photon propagation we only need to determine the metric components $h_{00}$, $h_{10}$, and $h_{11}$, evaluated at $\vec{x} = (x,0,0)$.
Furthermore, we need $\delta u^0_{S,D}$ and $\delta u^1_{S,D}$ in order to compute the Doppler shift (second line of \cref{eq:fshift}), again only along the $x^1$ axis. For free-falling $S$ and $D$, this requires solving the geodesic equation,
\begin{align}
    \frac{du^\mu}{d\tau} = - \Gamma^\mu_{\rho\sigma} u^\rho u^\sigma
                         = -\Gamma^\mu_{00} + \mathcal{O}(h^2) \,,
    \label{eq:du-dtau}
\end{align}
with
\begin{align}
  \Gamma^0_{00} = -\frac12 h_{00,0} \,,
  \qquad
  \Gamma^1_{00} = h_{01,0} - \frac12h_{00,1}\,,
  \label{eq:Christoffel}
\end{align}
and the other $\Gamma^i_{00} = 0$. (Note that here $\tau$ refers to the eigentime, not the measurement duration as in the main 
text). Let us remark that, alternatively, one could also use the equation of {\it geodesic deviation} to determine 
$\delta u^\mu(\lambda_S)-\delta u^\mu(\lambda_D)$; in the situation we consider here, however, 
we checked that this requires to consider this equation beyond the linear order in which it is typically stated.

The definition of the PD frame implies that at the origin, which we take to be the location of the photon source $S$, the spacetime metric remains locally flat even in the presence of GWs.
To second order in $x^i$, the metric is thus fully determined by the Riemann tensor evaluated at the origin.
As in linearized gravity the Riemann tensor is the same in all frames, we can conveniently express the metric in the PD frame in terms of the metric perturbation $h_{\mu\nu}^{TT}$ in the TT frame, where it has a particularly simple form.
This is true to {\it all} orders in $x^i$~\cite{Rakhmanov:2014noa,Berlin:2021txa,Domcke:2022rgu}.
For the class of experimental setups studied in this paper, we find that the relevant metric components in the PD frame take the following form:
\begin{align}
    h^{\rm PD}_{00} \big|_{\vec{x}=(x,0,0)}
        &= \frac{1 - \cos(x \omega_g c_\vartheta)}{\omega_g^2 c^2_\vartheta} \,
           (\partial_0)^2 h^{\rm TT}_{11} \big|_{\vec{x}=(0,0,0)}\,,
     \label{eq:h00PD}\\
    h^{\rm PD}_{01} \big|_{\vec{x}=(x,0,0)}
        &= h^{\rm PD}_{11} \big|_{\vec{x}=(x,0,0)} = 0  \,.
    \label{eq:h01PD}   
\end{align}
Using \cref{eq:htt11} for the metric perturbation in the TT frame, we can now calculate the contribution to the frequency shift that arises during photon propagation as
\begin{align}
    \frac{\omega_\gamma^D - \omega_\gamma^S}{\omega_\gamma^D} \bigg|_\text{prop}
        &= -\frac{\omega_0}{2} \int_0^{\lambda_D} \!\! d\lambda' \,
            \partial_0 \big[ h_{00}^{\rm PD}
              \big]_{x^\mu=(\lambda'\omega_0,\lambda'\omega_0,0,0)} \nonumber\\
        &= h_+ c_{\vartheta/2}^2 \Big\{
               \cos\varphi_0
             + t_\vartheta^2 \cos[\omega_g L + \varphi_0] \nonumber\\
        &\qquad\qquad~
             - c_\vartheta^{-2} \cos[\omega_g L (1 - c_\vartheta) + \varphi_0]
             \Big\} \nonumber \\
        &\quad
             + h_+ \omega_g L \frac{s_\vartheta t_\vartheta}{2} \sin(\omega_g L+\varphi_0) \,,
    \label{eq:photonprop}
\end{align}
with $t_\vartheta \equiv \tan\vartheta$.
For a rigid detector setup, this gives the observed frequency shift.

If on the contrary source and detector are free-falling, 
we need to include the contribution from the Doppler shift which follows from integrating
\begin{align}
    \frac{1}{2} \int_{\tau(\lambda=0)}^{\tau(\lambda=\lambda_D)} \! d\tau
        \big[ h^\text{PD}_{00,0} - h^\text{PD}_{00,1} 
        \big]_{x^\mu=(t(\tau),L,0,0)} \,.
    \label{eq:tau-integration}
\end{align}
However, naively using \cref{eq:htt11,eq:h00PD} in this expressions
does \emph{not} result in the same total frequency shift as the one derived in the TT frame, namely \cref{eq:fshift_freefall} in the main text. 
The reason for this discrepancy is a mismatch in our (implicit) choice of initial conditions: while the formulation in the TT frame, \cref{eq:htt11}, effectively assumes that the initial displacement of $D$ (relative to $S$) depends on the GW phase $\varphi_0$ at $t=0$, the formulation in the PD frame assumes this displacement to be zero at $t=0$, independently of $\varphi_0$.
One way to resolve this mismatch is to consider a GW that is only gradually switched on, reaching its full strength only after some finite time $t_0$.
This can formally be implemented by replacing \cref{eq:htt11} with
\begin{align}
    h^{\rm TT}_{11}(x^\mu)
        &= h_+ s^2_\vartheta \, \cos\!\left[ \omega_g (t - c_\vartheta x^1 - 
                                           s_\vartheta x^3) + \varphi_0 \right]
                                    \nonumber\\[0.2cm]
        &\quad
           \times \exp\big[- \epsilon (t_0 - t) \, \theta(t_0-t) \big] \,,
    \label{eq:htt11_adiab}
\end{align}
where $\epsilon > 0$ parameterizes the speed of the slow turn-on of the wave, and $\theta$ is the Heaviside function.
When taking the limit $\epsilon \to 0$, in particular, this clearly reproduces the plane wave solution adopted from \cref{eq:htt11}.
For any $\epsilon > 0$, on the other hand, this ansatz restores the Minkowski metric in the limit $t \to -\infty$, thereby uniquely fixing the initial conditions (i.e., both observers at rest) in a way that is independent of $t_0$. 

With this improved ansatz for $h^{\rm TT}$, and noting that $\tau = t$ to leading order in $h$, the Doppler shift in the PD frame becomes
\begin{align}
    \frac{\omega_\gamma^D - \omega_\gamma^S}{\omega_\gamma^D} \bigg|_\text{Doppler}
        \!\!
        &= \lim_{\epsilon \to 0} \frac{1}{2} \int_{-\infty}^L \!\! dt
           \big[h^{\rm PD}_{00,0} - h^{\rm PD}_{00,1} \big]_{x^\mu=(t,L,0,0)}
           \nonumber \\
        &= h_+ c_{\frac{\vartheta}{2}}^2 t_\vartheta^{2} \Big\{
               - \cos[\omega_g L + \varphi_0]            \nonumber\\
        &\qquad\qquad\quad~
               + \cos[\omega_g L (1 - c_\vartheta) + \varphi_0] \Big\}
                                                         \nonumber \\
        &\quad
         - h_+ \, \omega_g L \frac{s_\vartheta t_\vartheta}{2}
           \sin(\omega_g L+\varphi_0) \,.
    \label{eq:doppler_final}
\end{align}
Adding \cref{eq:photonprop,eq:doppler_final} we recover the TT frame result for the total frequency shift, \cref{eq:fshift_freefall}, as expected.
In the limit $\omega_g L \ll 1$ \cref{eq:doppler_final} and consequently \cref{eq:fshift_freefall} scales linearly with $\omega_g L$, despite the quadratic scaling of $h_{\mu\nu}^\text{PD}$ in \cref{eq:h00PD}. The reason for this is that the frequency shift $\Delta \omega_\gamma(t)$ is not simply proportional to $h_{\mu\nu}^\text{PD}(t)$, but instead required solving the geodesic equation over the duration of the photon travel time.
In particular, the $x^1$-derivative acting on $h_{00}^\text{PD}$ does not simply lead to an overall $\omega_g$ prefactor that would be cancelled by the $d\tau$ integration from \cref{eq:tau-integration}.

\section{Propagation of an electromagnetic wave in curved space time}
\label{app:MW}

In the absence of external currents, the inhomogeneous Maxwell equation in a general metric reads~\cite{landau1975classical}
\begin{align}
 \nabla_\mu ( g^{\mu \alpha} F_{\alpha \beta} g^{\beta \nu}) = \partial_\mu (\sqrt{-g} \; g^{\mu \alpha} F_{\alpha \beta} g^{\beta \nu}) = 0 .
\end{align}
Expanding to linear order in the GW amplitude 
yields
\begin{align}
 \partial_\mu F^{\mu \nu} + \partial_\mu (\tfrac{1}{2} h F^{\mu \nu} - h^{\mu \alpha} F_{\alpha}^{\; \nu} -  F^{\mu}_{\; \beta} h^{\beta \nu}) = 0 ,
\end{align}
where $F_{\mu \nu} = \partial_\mu A_\nu - \partial_\nu A_\mu$, $h \equiv h^\mu_\mu$ and we raise and lower indices with $\eta_{\mu \nu}$. We can now expand $F_{\mu\nu} = F^0_{\mu\nu} + \Delta F_{\mu\nu}$ with $F^0_{\mu\nu}$ solving Maxwell's equations in flat space, $\partial_\mu F_0^{\mu\nu} = 0$. With $\Delta F_{\mu\nu} = {\cal O}(h)$, this yields to linear order in $h$,
\begin{align}
 \partial_\mu \Delta F^{\mu \nu} = -   \partial_\mu ( \tfrac{1}{2} h F_0^{\mu \nu} - h^{\mu \alpha}  (F_0)_{\alpha}^{\; \nu} -  (F_0)^{\mu}_{\; \beta} h^{\beta \nu}) \,,
 \label{eq:MW}
\end{align}
i.e.\ the GW, together with the background EM field, acts as an effective current sourcing the induced EM field $\Delta F_{\mu \nu}$. 
The homogeneous Maxwell equation is not affected by the GW. 

Consider a linearly polarized EM wave with frequency $\omega_\gamma$ propagating along the $x$-axis,\footnote{
Choosing the orthogonal linear polarization (i.e.\ transverse to the plane spanned by the EM and GW wave vectors) gives similar results.
} 
\begin{align}
 \vec A(t,x) =  A_0 \cos[\omega_\gamma (t - x)]  \hat e_z + \vec{\Delta A}(t,x)  \,.
 \label{eq:Aansatz}
\end{align}
We use temporal gauge for the EM wave and the TT gauge for the GW. 
Physically, we expect small frequency shifts to appear as sidebands in Fourier space. Formally, we note that 
the right hand side of eq.~\eqref{eq:MW} can be expressed as a linear combination of $\cos_+(t,x)$ and $\cos_-(t,x)$ with
\begin{align}
 \cos_- & = \cos[(\omega_g - \omega_\gamma) t - (\omega_g c_\vartheta - \omega_\gamma) x ] \,, \nonumber \\ 
 \cos_+ & = \cos[(\omega_g + \omega_\gamma) t - (\omega_g c_\vartheta + \omega_\gamma) x] \,.
\end{align}
We thus make the ansatz
\begin{align}
 \vec{\Delta A}_i = a^-_i \cos_- + \, a^+_i \cos_+ \,.
 \label{eq:ansatz-part}
\end{align}
This solves eq.~\eqref{eq:MW} for
\begin{align}
 a_x^\pm & = A_0 h_+ s_\vartheta \frac{(\omega_g \pm \omega_\gamma(1 - c_\vartheta))}{4 (\omega_g \pm \omega_\gamma)} \nonumber  \\
 & = \frac{1}{4} A_0 h^+ s_\vartheta (1 - c_\vartheta) + {\cal O}(\omega_g/\omega_\gamma) \,, \\
 a_y^\pm & =  \tfrac{1}{4} A_0 h_\times  \,, \\
 a_z^\pm & = - A_0 h_+ \frac{\omega_\gamma}{\omega_g} \left( 8 \omega_\gamma (\omega_g \pm \omega_\gamma) \right)^{-1} \nonumber \\
 & \cdot \left( 2  c_\vartheta(\omega_g^2 \pm \omega_g \omega_\gamma + \omega_\gamma^2) \pm \omega_\gamma(\omega_g \pm 2 \omega_\gamma - \omega_g c_{2 \vartheta}) \right)\nonumber \\ 
 &= \mp \frac{1}{4} A_0 h^+ \frac{\omega_\gamma}{\omega_g} (1 + c_\vartheta) + {\cal O}[(\omega_g/\omega_\gamma)^0] \,. 
 \label{eq:az0}
\end{align}
Only the $z$-component features an $\omega_\gamma/\omega$ enhancement factor and we will therefore focus on this component in the following. 

From the above particular solution to eq.~\eqref{eq:MW} we obtain the general solution by adding the general solution of the homogeneous equation (i.e.\ the unperturbed Maxwell equation), which are simply regular plane waves:
\begin{align}
 A_z(t,x) = &  \tilde A_0 \cos[\omega_\gamma (t - x)] 
  + a_z^+ \cos_+ + a_z^- \cos_- \nonumber \\
 & + \sum_\lambda \alpha_z^\lambda \cos[(\omega_\gamma + \lambda \omega_g)( t - x)] \,,
 \label{eq:solution}
\end{align}
with $\lambda = \pm 1$. We now impose the boundary condition
\begin{align}
 A_z(t,0) = A_0 \cos(\omega_\gamma t) \,,
\end{align}
which yields
\begin{align}
 \tilde A_0 & = A_0  \\
 \alpha_z^\pm  & = \pm \frac{A_0 h^+ \omega_\gamma}{4 \omega_g} (1 + c_\vartheta) \,.
\end{align}
Arbitrary initial phases for the photon and GW can be recovered by substituting $\omega_\gamma t \mapsto \omega_\gamma t + \phi_\gamma$ and $\omega_g t \mapsto \omega_g t + \phi_g$ in the arguments of the cosines, respectively.
This directly reproduces all but the last term of eq.~\eqref{eq:A} in the main text. After inserting into the eom~\eqref{eq:MW} this term yields contributions of order $\omega_g/\omega_\gamma$, which we have dropped throughout this derivation. In conclusion, directly solving Maxwell's equation~\eqref{eq:MW} in curved space time fully reproduces the results of the more heuristic approach taken in the main text. The results of this appendix moreover confirm and extend the results found in Ref.~\cite{Melissinos:2010zz} for the special case of $\vartheta = \pi/2$.

\section{Long integration times}

\begin{figure}
    \centering
    \includegraphics[width=\columnwidth]{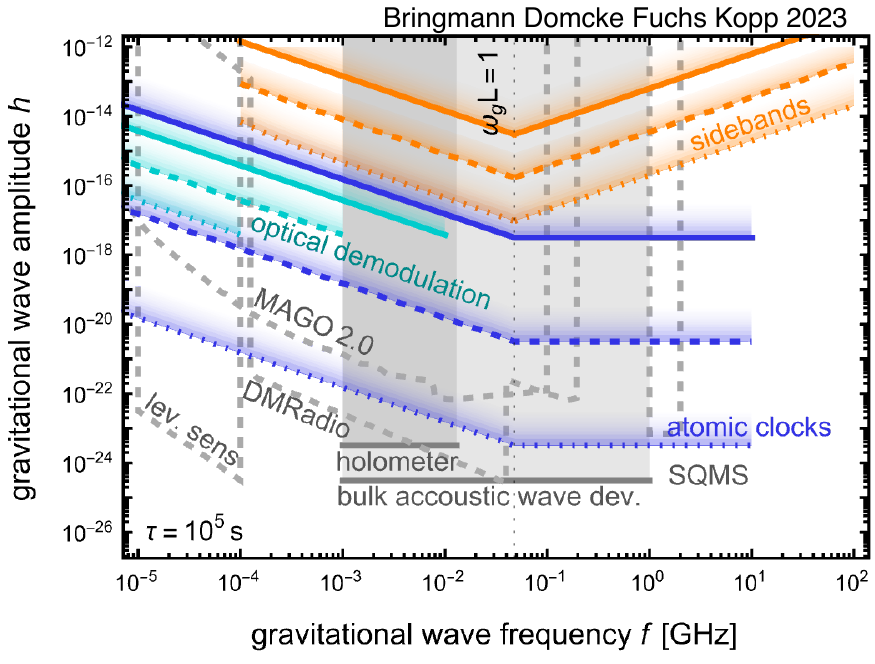}
    \caption{Sensitivity estimates for the three novel high-frequency GW detection methods presented here,
    in comparison to existing limits and proposed searches. Like \cref{fig:sens} in the main text, but for an integration time of 
    $\tau=10^5$\,s.
    }
    \label{fig:sens2}
\end{figure}

In the main text, we demonstrated the competitiveness of our newly proposed methods to detect high-frequency
GWs by comparing in \cref{fig:sens} the projected sensitivities to current and planned
alternative attempts to observe GWs in this frequency range. This comparison was based on 
an assumed integration time of $\tau=1$\,s, which in particular assumes the GW signal to last at least this long. Here, we complement this, for comparison, with the situation
where a much longer integration time of $\tau=10^5$\,s is adopted. The result is shown in \cref{fig:sens2}.
In terms of {\it relative} sensitivities, we thus find qualitatively the same as for short integration times.

\section{High-frequency GW sources}

Let us have a brief glance at possible GW signals in the frequency range considered here (see~\cite{Aggarwal:2020olq} for a detailed review).
For existing GW interferometers, the key target are binaries of black holes or other compact objects.
On the contrary, in the ultra high frequency regime above about \SI{10}{kHz}, there are no known astrophysical objects which are compact and small enough to lead to sizeable signals (see, though, 
ref.~\cite{Casalderrey-Solana:2022rrn} and discussion below).
However, exotic compact objects such as primordial black holes (PBHs) do generate signals in this range.
The strain amplitude at frequency $f = \omega_g / 2\pi$ for an inspiral signal from two PBHs with equal mass $m_\text{PBH}$ at distance $D$ is given by~\cite{Maggiore:2007ulw}
\begin{align}
    h \simeq 10^{-23}
             \bigg( \frac{\SI{10}{kpc}}{d} \bigg)
             \bigg( \frac{m_\text{PBH}}{\SI{e-5}{M_\odot}} \bigg)^{5/3}
             \bigg( \frac{f}{\SI{100}{MHz}} \bigg)^{2/3} \,.
\end{align}
The signal will persist until the black holes merge after
\begin{align}
    \tau \simeq \SI{0.02}{sec}
                \bigg( \frac{\si{MHz}}{f} \bigg)^{8/3}
                \bigg( \frac{\SI{e-5}{M_\odot}}{m_\text{PBH}} \bigg)  \,.
\end{align}
Such short transient signals constitute a serious challenge for high-frequency GW detectors.
Concretely, at $f = \SI{1}{MHz}$, only PBH binaries with $m_\text{PBH} < \SI{e-6}{M_\odot}$ will generate signals that last $\mathcal{O}(\SI{1}{sec})$.
Yet even if the PBH binary is located only \SI{1}{pc} away from the observer, the signal strain will be only $h \sim 10^{-22}$.
Obtaining a signal of similar strain but with a duration of \SI{e5}{sec} requires PBH as light as $\SI{e-9}{M_\odot}$ within \SI{1}{AU}.
For comparison, even if PBH make up all the dark matter in the Universe, their local density only will be about \num{8000}/pc$^3$ at $m_\text{PBH} = \SI{e-6}{M_\odot}$. Therefore, the probability that two PBH in our local neighborhood merge during our lifetime is exceedingly small.
Clearly, with the strain sensitivities currently achievable, detection of PBH signals is unlikely.
Nevertheless, PBH binaries provide a useful benchmark model and a target for future developments in this field.
In passing, we note that cosmological signals (i.e.\ stochastic GW backgrounds) are even more challenging to detect, due to stringent bounds on the total radiation energy density in the Universe from Big Bang nucleosynthesis and the cosmic microwave background.

Recently, the authors of ref.~\cite{Casalderrey-Solana:2022rrn} have made the very interesting observation that neutron star mergers could be a promising source of MHz gravitational waves. In particular, if quantum chromodynamics (QCD) undergoes a first-order phase transition at high baryon density, this phase transition could be triggered when the nuclear matter inside a neutron star is heated up and compressed during a merger. Whether or not such a phase transition exists is unknown, but if it exists, the dynamics of bubble nucleation and collision that characterize first-order phase transitions, as well as the resulting sound waves and turbulence, can lead to strong gravitational wave emission, with strains that may become detectable in the future.  This novel source of MHz gravitational waves is significant as it depends entirely on Standard Model dynamics, without requiring any ``new physics''.

\section{Optical Rectifiers}

\begin{figure*}
    \centering
    \includegraphics[width=\textwidth]{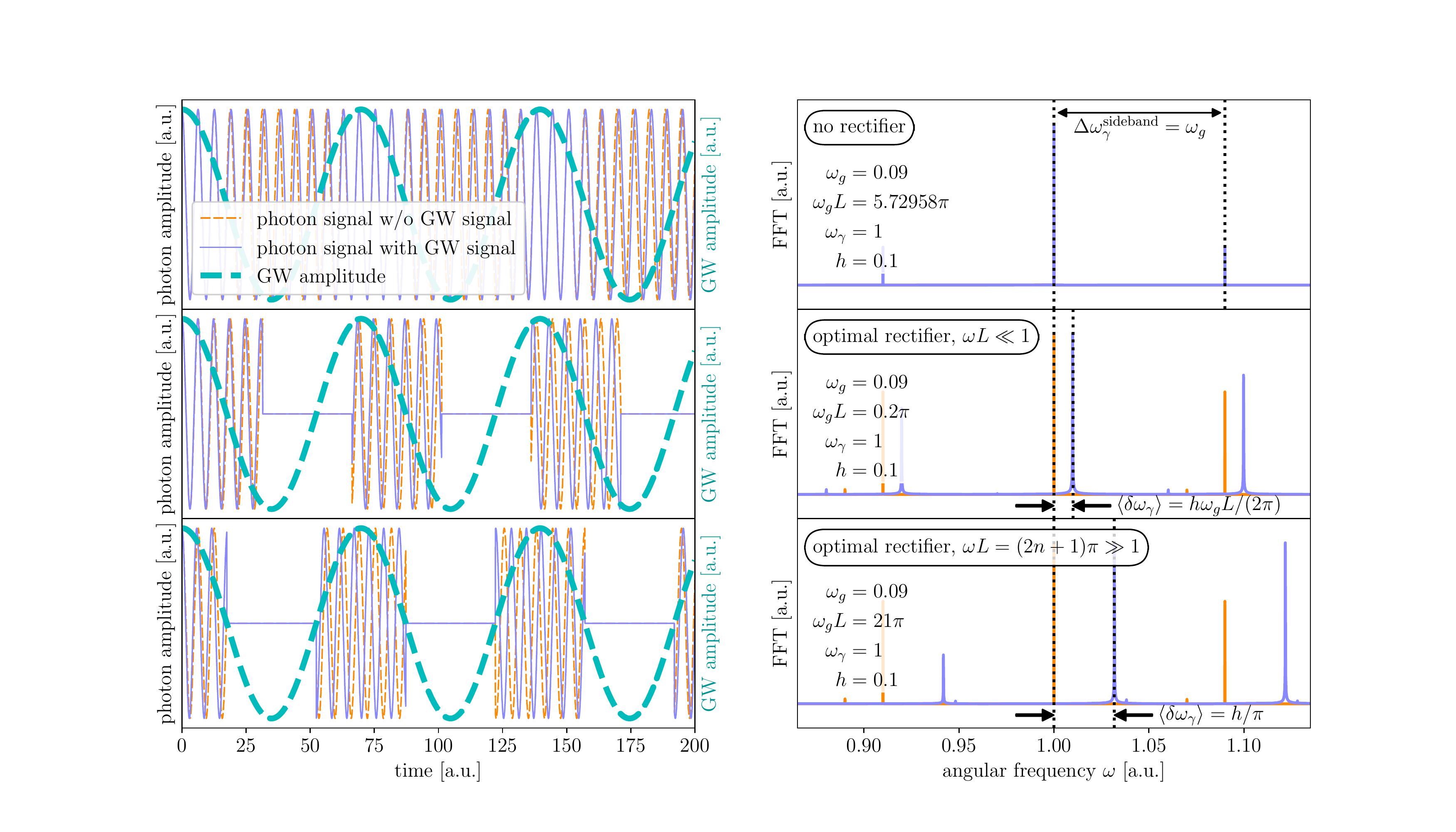}
    \caption{Effect of an optical rectifier on a photon signal perturbed by a gravitational wave. For a toy scenario in which the relevant effects are magnified by many orders of magnitude, we show on the left-hand side the photon amplitude without (orange dashed) and with (blue solid) the gravitational wave-induced perturbation, as well as the gravitational wave amplitude (thick cyan dashed).  The panels on the right show the Fourier transform of the photon signal.  We compare three cases: \textit{top:} uninterrupted photon signal (not rectified); in this case, the net frequency shift is zero, but we observe the emergence of faint sidebands. \textit{middle:} with an optical rectifier optimized for low-frequency gravitational waves ($\omega_g L \ll 1$), the carrier mode frequency is shifted, but the shift is suppressed by $\omega_g L$. \textit{bottom:} if $\omega_g L$ is a large odd integer multiple of $\pi$, much larger shifts are possible if both the gravitational wave frequency and its phase match the parameters that the rectifier is optimized for. Throughout, we have chosen $\vartheta = \pi/2$, i.e., a gravitational wave propagating orthogonal to the photons.}
    \label{fig:rectifier}
\end{figure*}

In the following, we provide further details on the proposed ``optical rectifiers'' that mask the photon signal during part of each GW cycle in such a way that a non-zero net frequency shift can be observed.  The effect of such a rectifier is illustrated in \cref{fig:rectifier}. Considering first the case \emph{without} rectifier (top panels), the figure confirms the results from the main text of this paper, namely the appearance of faint sidebands, separated from the carrier mode by $\Delta\omega_\gamma^\text{sideband} = \omega_g$.

With an optical rectifier chopping the photon signal into pieces, however, also the carrier frequency is shifted, see middle and bottom panels of \cref{fig:rectifier}, an effect that could be detectable using atomic clock techniques, as discussed in the main text. Note that the sidebands that are clearly visible in the middle and bottom panels reflect the action of the optical rectifier and are \emph{not} a means to detect the gravitational wave, as can be seen from the fact that they are present even in absence of a gravitational wave (orange dashed curves).

The amount by which the carrier mode is shifted, $\ev{\delta\omega_\gamma}$, can be understood by rewriting \cref{eq:fshift_freefall} as
\begin{align}
    \frac{\omega_\gamma^D - \omega_\gamma^S}{\omega_\gamma^D}
        &= h_+ c^2_{\vartheta\!/\!2}             
        \Big\{
                 \cos \varphi_0- \cos\!\big[\omega_g L (1 \!-\! c_\vartheta) + \varphi_0 \big]
            \Big\} \notag\\
        &= 2 h_+ c^2_{\vartheta\!/\!2}
           \sin\big[ \varphi_0 + \tfrac{1}{2} \omega_g L \big]
           \sin\big[ \tfrac{1}{2} \omega_g L \big] \,.
    \label{eq:fshift_freefall-appendix}
\end{align}
In the limit $\omega_g L \ll 1$ (middle panels of \cref{fig:rectifier}), $\tfrac{1}{2} \omega_g L$ can be neglected in the first sine, while the second sine can be expanded to linear order. An optimal rectifier, which we have assumed here, will allow photons to pass while $\sin\varphi_0 = \sin \omega_g t > 0$ and suppress the signal otherwise. Taking the time average in this case leads to the average photon frequency shift indicated in the plot, $\ev{\delta\omega_\gamma} = h \omega_g L / (2\pi)$ (for $\vartheta = \pi/2$ and `$+$' polarization, which we have assumed here).

If $\omega_g L \gg 1$, optimal sensitivity is achieved if $\omega_g L$ is an odd integer multiple of $\pi$. In this case, the second sine in \cref{eq:fshift_freefall-appendix} is unity, and an optimal optical rectifier will now allow photons to pass when $\sin[\varphi_0 + \pi/2] > 0$, as illustrated in the bottom panels of \cref{fig:rectifier}. The average frequency shift is then $\ev{\delta\omega_\gamma} = h / \pi$.

\section{How not to build a high-frequency GW detector}

Over the past years, several new detector concepts for high frequency GWs have been proposed, see ref.~\cite{Aggarwal:2020olq} for a review.
At the moment, which of these concepts (if any) can provide a viable way forward to reach the sensitivities needed to explore astrophysical and cosmological models remains an open and challenging question.
In view of this, we consider it worthwhile sharing some lessons learned about some of the challenges and difficulties in high-frequency GW detection, even though some of these may seem somewhat obvious in hindsight.
As in the main text of this paper, we will (largely) focus on GW detection through high-precision frequency measurements, but we expect many of the fundamental difficulties to be common among different detector concepts.

\paragraph*{Precision atomic measurements -- general considerations.} 
Frequency measurements up to and exceeding a relative accuracy of $10^{-20}$ have recently been achieved in a variety of different atomic systems, notably optical lattice clocks~\cite{Zheng:2021rnx, Bothwell:2021fqe}.
If no other suppression factors are at play, one might hope to translate this to dimensionless strain sensitivity for GW detection of the same order of magnitude.

However, three general obstacles prevent a direct exploitation of this sensitivity to GW searches. 

(i) The typical diameter of an atomic system is of order $L \simeq \SI{0.1}{nm}$, and thus much smaller than the wave length of $\mathcal{O}(\SI{100}{GHz})$ GWs, which is of order \si{mm}.
This implies that the signal is suppressed by a factor $\sim \omega_g L$, so that the GW amplitude cannot be fully exploited.
We are thus lead to focus on macroscopic systems, or alternatively the comparison of observables in atomic systems at macroscopically separated locations. 

(ii) While the mirrors in an interferometer are free-falling (or in the case of LIGO, VIRGO, and KAGRA~\cite{LIGOScientific:2014pky, VIRGO:2014yos, KAGRA:2020tym} suspended such that they are approximately free-falling in the direction of the GW force), atomic systems are tightly bound by electromagnetic forces. 
Thus any effects which rely on a direct competition of the force induced by the GW with electromagnetic binding forces are bound to be negligible. 
See~ref.~\cite{Ito:2020wxi} (and also refs.~\cite{Parker:1980kw, Leen:1983vu}) for a comprehensive summary of the interaction between GWs and an atomic system.
The combination of these two factors explains why even relatively large systems (in atomic units) such as Rydberg atoms with an atomic radius of ${\cal O}(\si{\mu m})$ provide only very limited sensitivity to GWs, with the leading order correction to the Coulomb potential, given by the GW induced correction to the gravitational potential, only leading to a relative energy shift of about $10^{-7} \, h \, (\omega_g/100~\text{MHz})^2$~\cite{Ito:2020wxi, Fischer:1994ed, Pinto:1995we, Siparov}.
In passing, note that this difficulty can under some circumstances become an advantage, as it implies that atomic systems can be viewed as spectrometers which are insensitive to GWs, thus providing an excellent tool to measure the gravitational red-shift (induced for example by a GW) experienced by photons travelling between different atomic systems located some distance apart. 

(iii) A crucial ingredient for the high precision quoted above is a long integration time which exceeds the inverse of the optical frequency by roughly the inverse of the obtained precision, that is, is at least of order seconds.
For high frequency GWs this implies that the waveform cannot be resolved.
The system thus averages over many GW oscillations, leading to a vanishing mean frequency shift.

Below, we give some concrete examples which suffer from these difficulties.
We emphasize that these examples do in no way prove that a given experimental technique is a priori unsuitable for high frequency GW detection, but rather that its naive application fails to achieve a competitive GW sensitivity.

\paragraph*{Atomic clocks.}
GW searches using frequency measurements by atomic clocks have been proposed in the mHz to \SI{10}{Hz} regime using atomic interferometers in space and on the ground~\cite{Graham:2017pmn, AEDGE:2019nxb, Badurina:2019hst, Alonso:2022oot}
or by placing atomic clocks on satellites~\cite{Loeb:2015ffa, Vutha:2015aza, Kolkowitz:2016wyg}.
The former proposal uses the slow velocity of the atoms to achieve a sensitivity to GWs of $\mathcal{O}(\si{Hz})$ frequency on a lab scale, the latter relies on a large distance of order $\omega_g^{-1}$ between the satellites.
In both cases, the relatively low frequency of the GW allows the resolution of its waveform.
As long as the GW frequency is below the optical frequency, atomic clocks at different locations are sensitive to the frequency shifts induced by the GW through photon propagation in the GW background as well as GW-induced Doppler shifts of the clocks, as discussed in the main text.
However, at frequencies above a \si{kHz}, the setup suffers from averaging effects, as described under (iii) above.
Hence, a direct measurement of the frequency shift becomes impossible, and one must resort to workarounds such as those described in the main text (side band measurements, optical demodulation, optical rectifier).

\paragraph*{M\"ossbauer spectroscopy.}
An alternative strategy to look for the tiny, GW-induced frequency shifts exposed by \cref{eq:fshift_freefall,eq:fshift_PD} is M\"ossbauer spectroscopy, which has first been considered in this context in ref.~\cite{Kaufmann:1970}.

M\"ossbauer spectroscopy exploits the fact that gamma radiation from atomic nuclei that form part of a solid body can be emitted recoil-free if the nucleus is tightly bound to its neighbours, such that the energy levels of the binding potential are far apart and ground state-to-ground state transitions have a large probability.
The recoil momentum is then absorbed by the whole macroscopic body rather than an individual nucleus.
M\"ossbauer emitters thus offer a source of nearly monochromatic gamma radiation which can, by the same arguments, be absorbed in a recoil-free manner by nuclei of the same isotope \cite{Moessbauer:1958, Frauenfelder:1962, Dyar:2006}.
Because both the emission and the absorption lines are very narrow, absorption is strongly resonantly enhanced.

A passing GW train will temporarily de-tune this resonance.
To observe this de-tuning, one could imagine a setup in which the gamma rays from a radioactive source pass through an absorber located a distance $L$ away, and the transmitted flux is measured by a detector located behind the absorber.
The sensitivity of this experiment will be inversely proportional to the achievable M\"ossbauer line width, $\Delta\nu$.
It should be possible to achieve effective relative line widths as small as $\Delta\nu/\nu \sim \num{e-18}$, notably in the isotope Ag-109m \cite{Coussement:1992a, Coussement:1992b}.
In fact, several groups have reported the observation of the M\"ossbauer effect in this transition \cite{Wildner:1979, Hoy:1988, Hoy:1990, Alpatov:2008, Bayukov:2009}, with one of them even claiming $\Delta\nu / \nu \sim \num{e-21}$ \cite{Alpatov:2008, Bayukov:2009}.

Unfortunately, this idea suffers from the same difficulties as the previous one: Narrow line widths require a long integration time, which leads to averaging over the GW signal.
Contrary to our proposals in the main text which utilize optical frequencies, these gamma rays can basically not be filtered, demodulated or rectified.

\paragraph*{Muon $g-2$ measurement.}
The anomalous magnetic moment of the muon, $g-2$, is among the most precisely measured quantities in particle physics.
To achieve this, the muon $g-2$ experiment at Fermilab~\cite{Muong-2:2021ojo} measures the precession frequency $\omega_s$ of muons on a circular orbit in an external magnetic field.
More precisely, the quantity of interest is $\omega_a = \omega_s - \omega_c \sim 10^{-2} \omega_c$, with $\omega_c \sim B/m_\mu$ denoting the cyclotron frequency at which muons orbit in the magnetic field.
The frequency difference $\omega_a$ can be measured with a relative accuracy of $10^{-6}$~\cite{Muong-2:2021ojo}.
This corresponds to a measurement of $\omega_c$ (or $\omega_s$) at the $10^{-8}$ level.
To estimate the impact of a passing GW on this measurements, we refer the reader to the formalism developed in ref.~\cite{Ito:2020wxi}, replacing the electron mass by the muon mass.
The advantage is that now we have a macroscopic experiment ${\cal O}(\SI{10}{m})$ in size, so we do not have to worry about $(\omega_g L)$ suppression.
Given that the muons are kept on their trajectory by external magnetic fields,  we expect the leading signal term to be the correction to the cyclotron frequency, ${\cal O}(h B /m_\mu)$.
This would allow the experiment to probe strain sensitivities of $h \sim 10^{-8}$, which is not particularly competitive.
In conclusion, despite the absence of penalizing suppression factors, in this case the accuracy of the measurement is simply insufficient to provide a good GW detector.

\bibliographystyle{JHEP_improved}
\bibliography{refs}

\end{document}